\documentclass[aps,prb,showpacs,tightenlines,twocolumn,nofootinbib,nobibnotes,superscriptaddress]{revtex4-1}
\usepackage{amsmath,amssymb,amsfonts,bm}
\usepackage{graphicx}
\usepackage{epstopdf}
\usepackage{dcolumn}
\usepackage{mathrsfs}
\usepackage{tgtermes}
\usepackage[colorlinks=true,linkcolor=blue,citecolor=blue, urlcolor=blue,bookmarks=false]{hyperref}

\bibpunct{[}{]}{,}{n}{}{}

\begin{document}
\title{Topological Anderson insulators in an Ammann-Beenker quasicrystal and a snub-square crystal}
\date{\today }
\author{Tan Peng}
\affiliation{Department of Physics, Hubei University, Wuhan 430062, China}
\author{Chun-Bo Hua}
\affiliation{Department of Physics, Hubei University, Wuhan 430062, China}
\author{Rui Chen}
\affiliation{Shenzhen Institute for Quantum Science and Engineering and Department of Physics, Southern University of Science and Technology, Shenzhen 518055, China}
\affiliation{School of Physics, Southeast University, Nanjing 211189, China}
\author{Dong-Hui Xu}
\affiliation{Department of Physics, Hubei University, Wuhan 430062, China}
\author{Bin Zhou}\email{binzhou@hubu.edu.cn}
\affiliation{Department of Physics, Hubei University, Wuhan 430062, China}

\begin{abstract}
The quest for the topological phases of matter in an aperiodic system has been greatly developed recently. Here we investigate the effects of disorder on topological phases of a two-dimensional Ammann-Beenker tiling quasicrystalline lattice. For comparison purposes, we also consider the case of a periodic snub-square crystalline lattice, which has the same primitive tiles as the Ammann-Beenker tiling quasicrystalline lattice. By calculating the spin Bott index and the two-terminal conductance, we confirm that the topological phases with disorder share the similar properties in the two systems which possess different symmetry and periodicity. It is shown that the quantum spin Hall states are robust against weak disorder in both the quasicrystalline lattice and the crystalline lattice. More interesting is that topological Anderson insulator phases induced by disorder appear in the two systems. Furthermore, the quantized conductance plateau contributed by the topological Anderson insulator phase is verified by the distribution of local currents.

\end{abstract}

\maketitle

\section{Introduction}
In the past decade, the study of quantum spin Hall (QSH) insulator, which is a two-dimensional (2D) topological insulator with a full insulating bulk gap but gapless helical edge states protected by time-reversal symmetry, has always been a hot spot in condensed matter physics \cite{
RevModPhys.82.3045,RevModPhys.82.1959,RevModPhys.83.1057,RevModPhys.88.021004,RevModPhys.89.041004,
PhysRevLett.95.226801,PhysRevLett.95.146802,bernevig2006quantum,PhysRevLett.96.106802,
PhysRevLett.97.036808}.
However, most of the previous works were implemented in the crystalline systems. The aperiodic systems without the translational symmetry are also found to host topological phases \cite{tanaka1987topological,PhysRevLett.118.236402,
PhysRevB.96.121405,PhysRevB.96.100202,mitchell2018amorphous,Bourne_2018,poyhonen2018amorphous,
PhysRevB.99.165413,Chern_2019,PhysRevLett.123.076401,PhysRevB.99.045307,PhysRevResearch.2.012067,
PhysRevB.91.085125,PhysRevX.6.011016,PhysRevLett.122.237601,
PhysRevB.101.020201,PhysRevLett.124.036803,PhysRevLett.123.196401,hua2020higherorder,
PhysRevLett.116.257002,ghadimi2020topological,PhysRevLett.121.126401,PhysRevB.98.125130,PhysRevB.100.085119,
PhysRevLett.109.106402,PhysRevB.91.064201,PhysRevB.88.125118,PhysRevB.100.214109,doi:10.1063/1.5083051,
PhysRevLett.119.215304,PhysRevX.9.021054,PhysRevResearch.2.033071,PhysRevLett.123.196401,Silva_2019,
PhysRevB.94.205437,PhysRevX.6.011016}. Interestingly, topological phases in quasicrystalline systems, which are the special aperiodic systems and possess forbidden symmetries in crystals, have been extensively studied.
Some typical topological phases, such as the non-Hermitian topological phase \cite{PhysRevLett.122.237601,
PhysRevB.101.020201}, the higher-order topological phase \cite{PhysRevLett.124.036803,PhysRevLett.123.196401,hua2020higherorder}, and the topological superconductor phase \cite{PhysRevLett.116.257002,ghadimi2020topological}, have been studied in quasicrystalline systems. Moreover, the photonic quasicrystals \cite{PhysRevLett.110.076403} and the quasiperiodic acoustic waveguides \cite{PhysRevLett.122.095501} can be used as the experimental platform to research topological phase in quasicrystals.

Very recently, Huang and Liu theoretically presented that the fivefold Penrose tiling quasicrystal can be used as a platform to realize the QSH insulator, and a real-space topological invariant called the spin Bott index is defined to characterize the topological phase of the system \cite{PhysRevLett.121.126401,PhysRevB.98.125130}. Furtherly, they found that the QSH states in an eightfold Ammann-Beenker tiling quasicrystal and a periodic snub-square crystal, both sharing the same primitive tiles, exhibit similar topological behaviors regardless of symmetry and periodicity, except for some quantitative differences \cite{PhysRevB.100.085119}.

However, due to the abundant physical phenomenon induced by disorder, the interplay between disorder and topology is an important topic in the research of topological phase. Topological Anderson insulator (TAI) \cite{PhysRevLett.102.136806}, which is a disorder-induced topological phase, has been theoretically studied in various systems, including Chern insulators \cite{PhysRevLett.105.115501,Zhang_2013,PhysRevB.92.085410,PhysRevLett.116.066401,PhysRevB.100.054108}, topological insulators \cite{PhysRevB.80.165316,PhysRevLett.103.196805,Wu_2016,PhysRevB.83.045114,
PhysRevB.91.214202,PhysRevB.96.205304,PhysRevB.84.035110,orth2016topological,PhysRevLett.105.216601,
PhysRevB.82.115122,PhysRevLett.113.046802}, topological semimetals \cite{PhysRevLett.115.246603,PhysRevB.93.075108,PhysRevB.95.245305,PhysRevB.97.235109,PhysRevB.98.235159,PhysRevB.97.024204,
PhysRevLett.116.066401}, topological superconductors \cite{PhysRevB.93.125133,qin2016disorder,PhysRevB.98.134507,PhysRevB.100.205302}, non-Hermitian systems \cite{PhysRevA.101.063612,zhang2020non,Liu_2020}, and higher-order topological insulators \cite{PhysRevLett.125.166801,yang2020higher}. The more important discoveries of TAI are the recent experimental realizations in a one-dimensional disordered atomic wires \cite{meier2018observation}, a photonic platform \cite{stutzer2018photonic,PhysRevLett.125.133603} and electric circuits \cite{zhang2020experimental}, respectively. It is noted that recently Chen \emph{et al.} extended the territory of the TAI phase from crystalline systems to quasicrystalline systems, and they implemented the TAI in a fivefold Penrose tiling quasicrystal \cite{PhysRevB.100.115311}. Inspired by the work of Huang and Liu \cite{PhysRevB.100.085119}, an intriguing question is that whether the appearance of TAI phase in crystalline and quasicrystalline systems is similar and has the same regular pattern.

\begin{figure}[tp]
	\includegraphics[width=8cm]{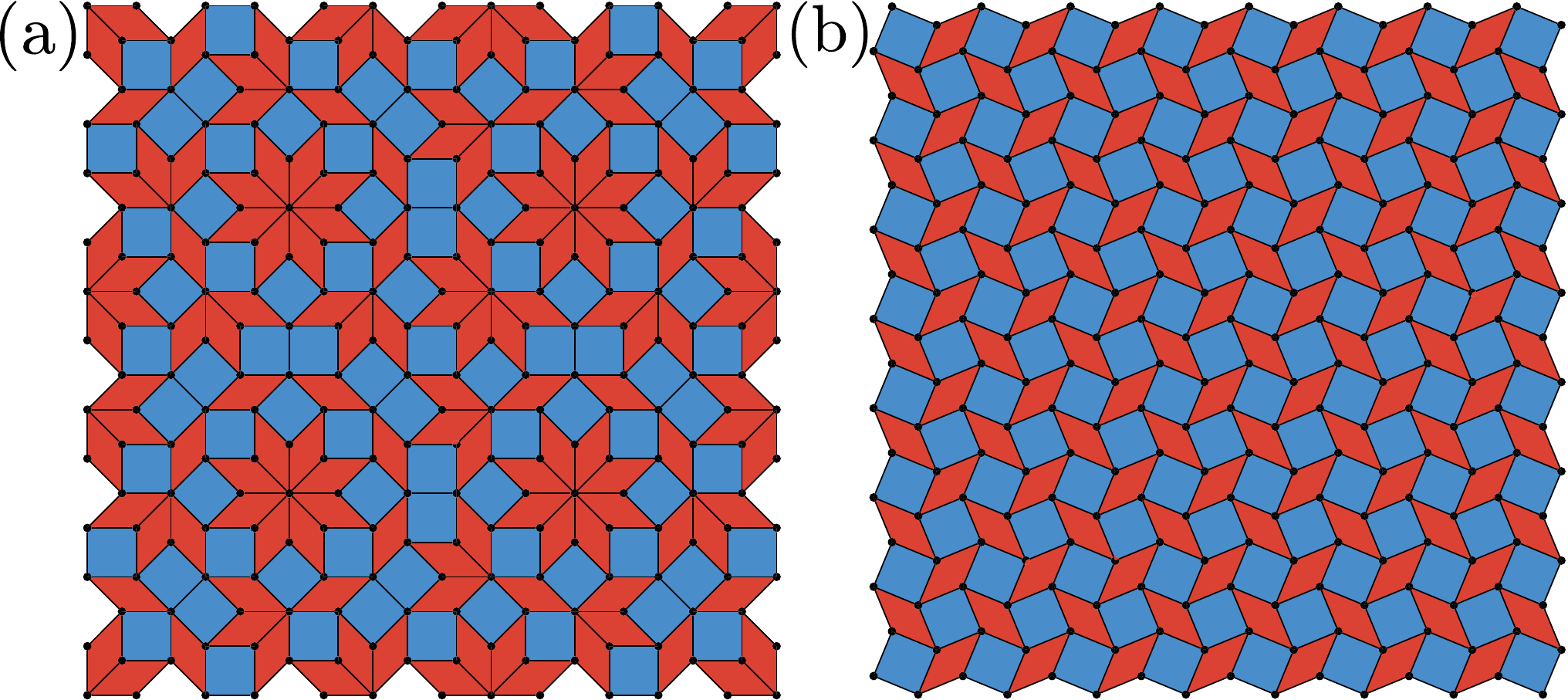} \caption{(a) Ammann-Beenker tiling quasicrystal containing $264$ vertices and (b) snub-square crystal containing $256$ vertices. The two lattices are consist of the same square and rhombus tiles. The first three nearest-neighbor bonds correspond to the short diagonal of the rhombus tile, the edge of square and rhombus tile, and the diagonal of square tile, respectively. The distance ratio of the three bonds is $r_{0}:r_{1}:r_{2}=2\sin \frac{\pi }{8}:1:2\sin \frac{\pi }{4}$.}%
\label{fig1}
\end{figure}

In this work, we investigate the effects of disorder on topological phases of a 2D eightfold Ammann-Beenker tiling quasicrystalline lattice \cite{grunbaum1987tilings,c641bbfabe714fefbd7c37e571cb29fa,Duneau_1989,kramer2003coverings} and a periodic snub-square crystalline lattice  \cite{chavey1989tilings}, as shown in Fig.~\ref{fig1}. This quasicrystalline lattice and this crystalline lattice have the same square and rhombus tiles, therefore they can be used as an effective comparison device to study the similarities and differences of physical phenomenon in the systems with different symmetry and periodicity \cite{PhysRevB.100.085119}. By calculating the spin Bott index and the two-terminal conductance, we show that the QSH insulator phases in the two systems are robust against weak disorder, and the TAI induced from a normal insulator occurs at a certain region of disorder strength with a quantized conductance plateau ($G={2e^2}/{h}$) characterized by a nonzero spin Bott index ($B_{s}=1$). More interestingly, we found that with the same model parameters, the disorder-induced topological phase transition can occur in smaller disorder strength in the quasicrystalline lattice than in the crystalline lattice. This is due to the different initial bulk energy gaps of the two systems. The distribution of nonequilibrium local current further confirms that the quantized conductance plateau arises from the helical edge states induced by disorder. It is presented that the topological nontrivial phases with disorder reveal qualitative similarities in quasicrystals and crystals, in spite of quantitative differences. Thus, it is indicated that the appearance of TAI is independent of the lattice symmetry and periodicity.

The rest of the paper is organized as follows. In Sec.~\ref{Model}, we introduce a QSH insulator model with Anderson-type disorder in 2D quasicrystalline lattice and crystalline lattice. Then, we give the details of numerical methods in Sec.~\ref{Methods}, and provide numerical results for studying the topological phase transitions of the two systems in Sec.~\ref{NS}. Finally, we summarize our conclusions in Sec.~\ref{Conclusion}.

\section{Model}
\label{Model}
We start with a tight-binding model of a QSH insulator with Anderson-type disorder in an Ammann-Beenker tiling quasicrystalline lattice and a periodic snub-square crystalline lattice. The lattice sites are located on the vertices of the quasicrystalline lattice and the crystalline lattice as shown in Fig.~\ref{fig1}. In this work, we only consider the first three nearest-neighbor hopping. The model Hamiltonian is given by \cite{PhysRevLett.124.036803,hua2020higherorder}
\begin{equation}
H=H_{0}+H_{\text{dis}},
\label{H}
\end{equation}
with the QSH insulator Hamiltonian
\begin{align}
H_{0}=&-\sum\limits_{j\neq k}\frac{l(r_{jk})}{2}c_{j}^{\dag}[it_{1}(s_{3}\tau _{1}\cos \theta _{jk}+s_{0}\tau _{2}\sin \theta_{jk})\nonumber  \\
&+t_{2}s_{0}\tau _{3}]c_{k}+\sum\limits_{j}(M+2t_{2})c_{j}^{\dagger}s_{0}\tau _{3}c_{j},
\label{H0}
\end{align}
and the Anderson-type disorder term
\begin{equation}
H_{\text{dis}}=\sum\limits_{j}c_{j}^{\dagger}U_{j}s_{0}\tau _{0}c_{j},
\label{Hdis}
\end{equation}
where  $c_{j}^{\dag }=(c_{j\alpha \uparrow }^{\dag },c_{j\alpha \downarrow }^{\dag
},c_{j\beta \uparrow }^{\dag },c_{j\beta \downarrow }^{\dag })$  is the creation  operator at site $j$. $\alpha$ and $\beta$ present two orbitals at each site. $\uparrow$ and $\downarrow$ denote spin-up and spin-down, respectively. $j$ and $k$ denote lattice sites running from $1$ to $Q$, and $Q$ is the total number of lattice sites. $s_{1,2,3}$ and $\tau_{1,2,3}$ are the Pauli matrices acting an the spin and orbital spaces, respectively. $s_{0}$ and $\tau_{0}$ are the $2\times 2$ identity matrices. $t_{1}$ and $t_{2}$ are the hopping amplitudes.  $\theta_{jk}$ is the polar angle of bond between site $j$ and $k$ with respect to the horizontal direction. $l(r_{jk})=e^{1-r_{jk}/\lambda }$ is the spatial decay factor of hopping amplitudes with the decay length $\lambda $, where $r_{jk}=|\mathbf{r}_{j}-\mathbf{r}_{k}|$ is the distance from the site $j$ to site $k$. $M$ is the Dirac mass. $U_{j}$ is the uniform random variable chosen from $[-W/2, W/2]$, and $W$ is the disorder strength. $H_{0}$ preserves time-reversal, particle-hole and chiral symmetries, therefore it belongs to the class DIII \cite{PhysRevB.55.1142,PhysRevB.78.195125,PhysRevLett.124.036803}. It is worth noting that the system will reduce to a Bernevig-Hughes-Zhang model \cite{bernevig2006quantum} which describes the QSH states in HgTe quantum wells, if the QSH insulator Hamiltonian in Eq. 2 is projected to a square lattice and only the nearest neighbor hopping is considered. In the following calculations, the spatial decay length $\lambda$ and the side length of rhombus and square $r_{1}$ are fixed as $1$, and the energy unit is set as $t_{1}=1$.

\section{Numerical Methods}
\label{Methods}
In this section, we introduce the numerical methods, which include the spin Bott index \cite{PhysRevLett.121.126401,PhysRevB.98.125130,PhysRevB.100.115311} and the two-terminal conductance based on the recursive Green's function method \cite{MacKinnon_1985,PhysRevB.72.235304}, to study the disorder-induced topological phase transitions in the quasicrystalline and the crystalline systems.

It is well known that the $\mathbb{Z}_{2}$ index is a useful topological invariant to characterize the QSH topological phase in periodic systems \cite{PhysRevLett.95.146802,fu2006time}. However, the Amman-Beenker tiling quasicrystal lacks the translational symmetry since it is tiled completely of the 2D plane with squares and rhombuses in an aperiodic way. Therefore, the quasicrystalline systems can not use the momentum-space $\mathbb{Z}_{2}$ index to characterize the topological phase, and we need to adopt a real-space topological invariant without translational invariance. It is worth noting that Huang and Liu have proposed that the spin Bott index, which is a real-space topological invariant, can appropriately characterize the QSH state in both periodic and aperiodic systems \cite{PhysRevLett.121.126401,PhysRevB.98.125130}. Additionally, there are several $\mathbb{Z}_{2}$ invariants that can be computed on 2D systems with time-reversal symmetry and without translational invariance in finite-dimensional Hilbert space \cite{li2019local,loring2011disordered}. Here, we will employ the spin Bott index to characterize the topological phases of the systems.

Here, we review the detailed steps of numerical calculation of the spin Bott index. First, one constructs the projector operator of the occupied states as
\begin{align}
 P=\sum_{i}^{N}|\psi _{i}\rangle \langle \psi _{i}|,
\label{P}
\end{align}
where $\psi _{i}$ is the $i$th wave function of the Hamiltonian (\ref{H}) and $N$ is the total number of occupied states. Then, one introduces another projector operator as $P_{z}=P\hat{\eta}_{z}P$, where $\hat{\eta}_{z}=\frac{\hbar }{2}\sigma _{z}$ is the spin operator with the Pauli matrix $\sigma _{z}$. The eigenvalues of $P_z$ are divided into two parts by zero energy, in which the number of positive and negative eigenvalues are both equal to $N/2$. Then, a new projector operator can be constructed as
\begin{align}
P_{\pm }=\sum_{i}^{N/2}\left\vert \phi _{i}^{\pm }\right\rangle \left\langle \phi _{i}^{\pm }\right\vert,
\label{P1}
\end{align}
where $\phi _{i}^{+}$ and $\phi _{i}^{-}$ are the eigenvectors corresponding to the $i$th positive and negative eigenvalues, respectively.
The projected position operators of the two spin sectors can be defined as
\begin{align}
&U_{\pm }=P_{\pm }e^{i2\pi X}P_{\pm }+(I-P_{\pm }),\\
&V_{\pm }=P_{\pm }e^{i2\pi Y}P_{\pm }+(I-P_{\pm }),
\label{UV}
\end{align}
where $X$ and $Y$ are two diagonal matrices, $X_{ii}=x_{i}$ and $Y_{ii}=y_{i}$ with $(x_{i},y_{i})$ being the coordinate of the $i$th lattice site.

In addition, the singular value decomposition (SVD) method, acting on $U_{\pm }$ and $V_{\pm }$, will be used to improve the stability of the numerical results. The SVD can be expressed as $S=Z\Lambda \Pi ^{\dag }$, where $Z$ and $\Pi$ are unitary and $\Lambda$ is real and diagonal. Then, one takes $\tilde{S}=Z\Pi ^{\dag }$ as the new projected position operator used to replace $S$. Finally, one can obtain the spin Bott index as \cite{PhysRevLett.121.126401,PhysRevB.98.125130,PhysRevB.100.115311}
\begin{align}
B_{s}=\frac{1}{2}(B_{+}-B_{-}),
\label{Bott}
\end{align}
where $B_{\pm }=\frac{1}{2\pi }{\rm Im}\{{\rm Tr}[\ln (\tilde{V}_{\pm }\tilde{U}_{\pm }\tilde{V}_{\pm }^{\dag
}\tilde{U}_{\pm }^{\dag })]\}$
are the Bott indexes of spin up and down, respectively. The case with $B_{s}=0$ corresponds to the normal insulator phase, and $B_{s}=1$ corresponds to the QSH insulator phase.

\begin{figure}[tp]
	\includegraphics[width=8cm]{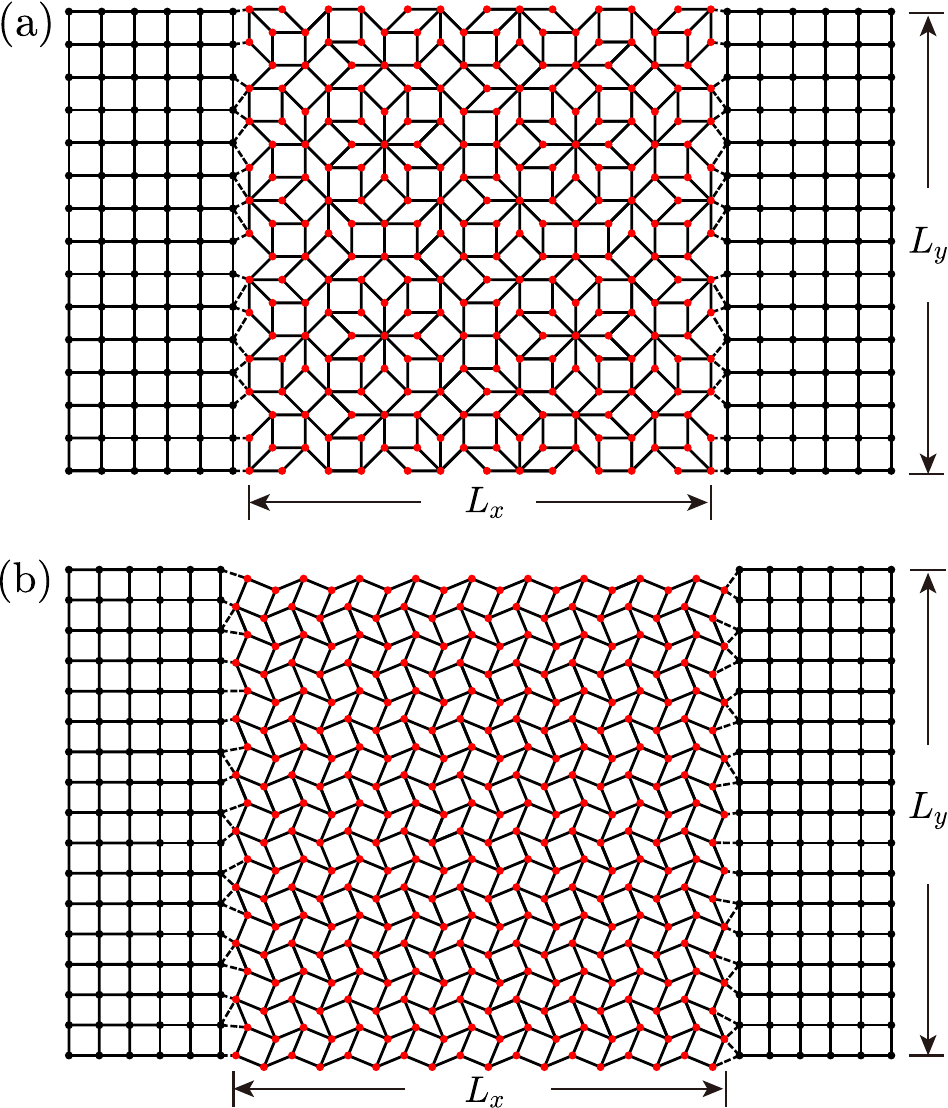} \caption{Schematic illustrations of the two-terminal setups of (a) the Ammann-Beenker tiling quasicrystal device and (b) the snub-square crystal device. The device size is $L_{x}\times L_{y}$, with $L_{x}=200r_1$ and $L_{y}=100r_1$, respectively.}%
\label{fig2}
\end{figure}

However, we also study the transport properties of the systems, and the setups are shown in Fig.~\ref{fig2}. It is assumed that two semi-infinite normal metal leads are attached to the end of the quasicrystal device [shown in Fig.~\ref{fig2}(a)] and the crystal device [shown in Fig.~\ref{fig2}(b)], respectively. Then, we calculate the two-terminal conductance of the system based on the recursive Green's function method \cite{MacKinnon_1985,PhysRevB.72.235304}. According to Landauer-B\"uttiker-Fisher-Lee formula \cite{Landauer_1970,B_ttiker_1988,Fisher_1981}, the conductance can be written as
\begin{align}
G=\frac{e^2}{h}T(\mu ),
\label{Conductance}
\end{align}
where $T(\mu )= \text{Tr}[\Gamma _{L}(\mu )G^{r}(\mu )\Gamma _{R}(\mu )G^{a}(\mu )]$ is the transmission coefficient at energy $\mu$. $\Gamma _{L(R)}(\mu)=i(\Sigma_{L(R)}^{r}-\Sigma_{L(R)}^{a})$ is the left (right) linewidth with the left (right) lead retarded self-energy $\Sigma^{r} _{L(R)}$ and the left (right) lead advanced self-energy $\Sigma^{a} _{L(R)}$. $G^{r(a)}(\mu)$ is the retarded (advanced) Green's function of the device, and can be expressed as
\begin{align}
G^{r}(\mu )=[G^{a}(\mu )]^{\dag }=[\mu -H-\Sigma _{L}^{r}-\Sigma_{R}^{a}]^{-1},
\label{r-a-Conductance}
\end{align}
where $H$ is the device Hamiltonian.

\section{Numerical simulation}
\label{NS}
In this section, we give the numerical results of calculating the spin Bott index and the two-terminal conductance to investigate the topological phase transitions of the quasicrystalline and the crystalline systems.

\begin{figure}[tp]
	\includegraphics[width=8.5cm]{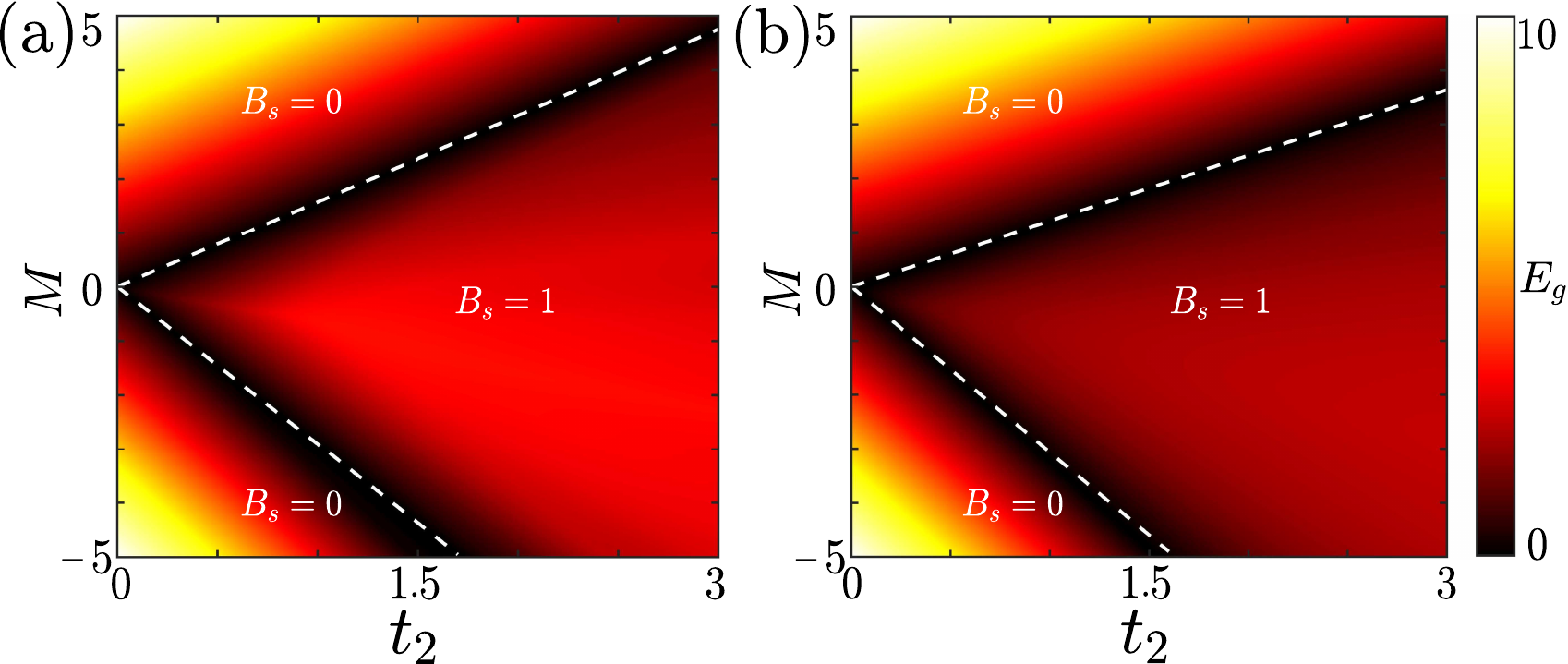} \caption{Topological phase diagrams of (a) the Ammann-Beenker tiling quasicrystal and (b) the snub-square crystal in the ($M$, $t_2$) space with $W=0$. The color map represents the magnitudes of the bulk energy gap. The white dashed lines represent the phase boundaries calculated by the spin Bott index, and the approximate linear equations are (a) $M/t_{2}=1.65$, $M/t_{2}=-2.91$ and (b) $M/t_{2}=1.21$, $M/t_{2}=-3.06$.}%
\label{fig3}
\end{figure}
\begin{figure*}[tp]
    \centering
	\includegraphics[width=\linewidth,scale=1.00]{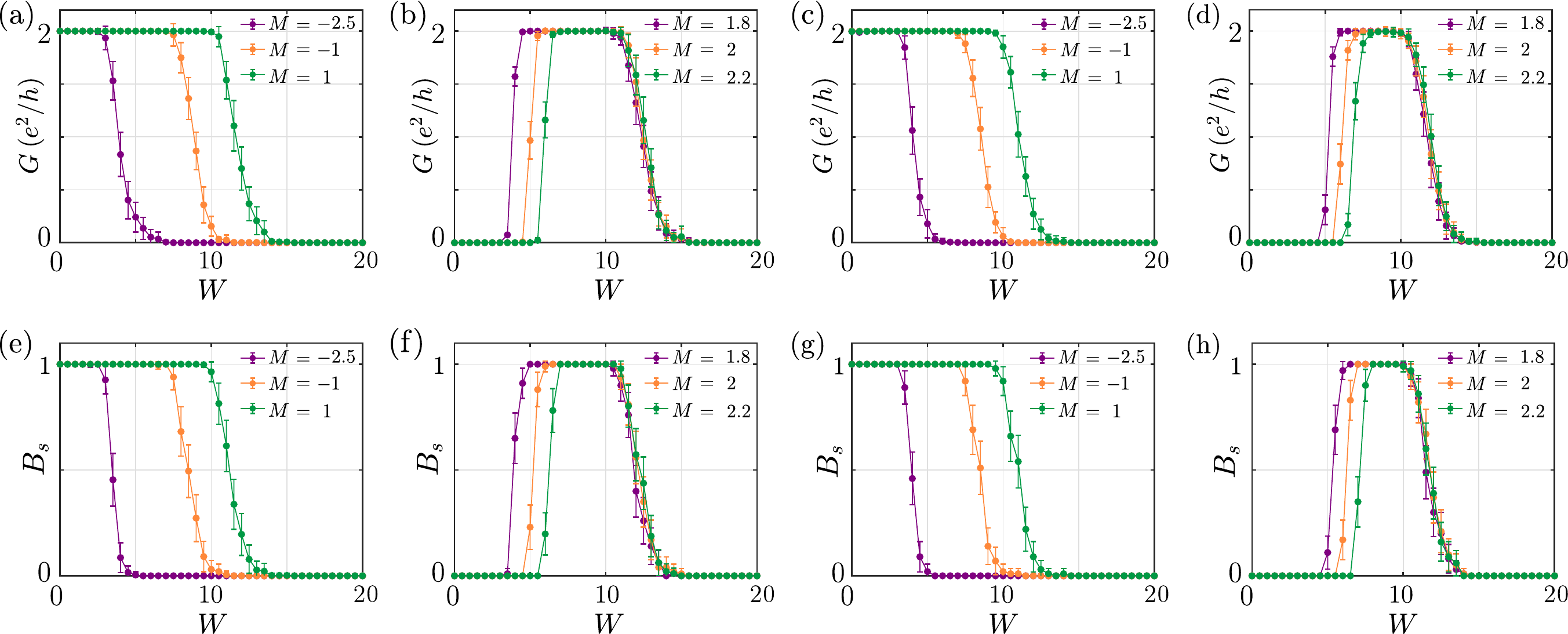} \caption{ The two-terminal conductance $G$ as a function of the disorder strength $W$ with different Dirac mass $M$ for (a,b) the quasicrystalline system and (c,d) the crystalline system. The chemical potential is taken as $\mu=0$. The spin Bott index $B_{s}$ as a function of $W$ with different Dirac mass $M$ for (e,f) the quasicrystalline system and (g,h) the crystalline system. The error bars show the standard deviation of the conductance for $1000$ samples. We set $t_2=1$. The conductance is calculated with the device size being $L_{x}\times L_{y}=200r_1\times100r_1$. The spin Bott index is calculated with a square sample containing 1393 sites for quasicrystalline system and 1444 sites for crystalline system, respectively.}%
\label{fig4}
\end{figure*}

Before studying disorder effects on the two systems, we first reveal the topological phase diagrams of the two systems in the clean limit, i.e., the disorder strength $W=0$. Figure~\ref{fig3} shows the bulk energy gap as a function of Dirac mass $M$ and hopping amplitude $t_2$ in the clean limit for the two systems. The color map shows the magnitudes of the bulk energy gap $E_{g}$. The energy gaps of two systems are both calculated numerically in real space with periodic boundary condition. There would be midgap states if we directly covert a square-shaped AB tiling quasicrystalline lattice to a torus geometry when applying periodic boundary conditions. This because some additional hoppings are introduced when linking the sites on the opposite boundaries. To avoid this, we first generate a square-shaped quasicrystalline lattice and then delete the sites on the top boundary of the quasicrystal square (these sites are equivalent to the sites on the opposite bottom boundary). Next, we glue the bottom and top boundaries. And we perform the same procedure for the right and left boundaries. With this method, no additional hoppings are introduced on the tortus geometry.

Both of the two phase diagrams are separated into three parts by two lines of energy gap closure, which indicates that the bulk energy gap of the two systems has a process of closing and reopening with the change of parameters. It also means that topological phase transitions may occur in the two systems. Therefore, we also compute the spin Bott index to distinguish the topological phases of the two systems, and give the phase boundaries of the normal insulator phase and the QSH insulator phase (the white dashed lines in Fig.~\ref{fig3}). In the case of quasicrystalline lattice, as shown in Fig.~\ref{fig3}(a), it is found that a QSH insulator phase is characterized by the nonzero spin Bott index $B_{s}=1$ when $-2.91<M/t_{2}<1.65$. When $M/t_{2}<-2.91$ and $M/t_{2}>1.65$, two normal insulator phases with $B_{s}=0$ are shown. Simultaneously, for the snub-square crystalline lattice, the phase boundaries are $M/t_{2}=1.21$ and $M/t_{2}=-3.06$ as shown in Fig.~\ref{fig3}(b). It is found that the phase diagrams of two systems are very similar expect that the slopes of the phase boundaries are slightly different in the clean limit by comparing Figs.~\ref{fig3}(a) and ~\ref{fig3}(b).

Next, we study the disorder effects on the topological phase transitions of the two systems. Figure~\ref{fig4} shows the two-terminal conductance and the spin Bott index as a function of the disorder strength $W$ at several specified parametric spatial points for two systems.

As shown in Fig.~\ref{fig3}(a), for the case of $M_{l}=(-2.5, -1, 1)$ in quasicrystalline system, the phases are the QSH insulator phase with the nonzero spin Bott index $B_{s}=1$ in the clean limit. The corresponding bulk energy gap are $E_{l}\approx(0.526, 3.136, 1.117)$, where $l=1, 2, 3$. With the disorder strength increasing, it is found that the QSH insulator phases remain stable in the case of weak disorder, which are characterized by the quantized conductance $G=2e^2/h$ [colourful curves in Fig.~\ref{fig4}(a)] and the nonzero spin Bott index $B_{s}=1$ [colourful curves in Fig.~\ref{fig4}(e)]. When the disorder strength $W$ reach the critical values, the system undergoes topological phase transitions from QSH phase into normal insulator phase with the conductance $G=0$ and the spin Bott index $B_{s}=0$. For the case of $M_{l}=(-2.5, -1, 1)$ in crystalline system, the numerical results of calculating the conductance and the spin Bott index, as shown in Figs.~\ref{fig4}(c) and~\ref{fig4}(g), are similar with the case of quasicrystalline lattice except that the critical values of the phase transition are slightly different. The numerical results clearly demonstrate that the topological insulator phase is robust against weak disorder whether in quasicrystal system or crystalline system.

For another case of $M_{l}=(1.8, 2, 2.2)$ in quasicrystalline system, the phases are the normal insulator phases with the spin Bott index $B_{s}=0$ in the clean limit, as shown in Fig.~\ref{fig3}(a). The corresponding bulk energy gap are $E_{l}\approx(0.511, 0.905, 1.299)$. The purple, orange and green curves in Fig.~\ref{fig4}(b) show the two-terminal conductance in the quasicrystalline system as a function of disorder strength with $M_{l}=(1.8, 2, 2.2)$, respectively. It is found that the conductances change from $G=0$ to $G=2e^2/h$ at $W_{l}\approx(5, 6, 7 )$ and return to $G=0$ at $W\approx10.5$. Some remarkable plateaux with the quantized conductance $G=2e^2/h$ are revealed. The quantized plateaux indicate that a QSH insulator phase induced by disorder in the quasicrystalline system appears, which is so called topological Anderson insulator \cite{PhysRevLett.102.136806,PhysRevB.100.115311}. Furthermore, the values of spin Bott index jump from $B_{s}=0$ to $B_{s}=1$ at $W_{l}\approx(5, 6, 7 )$ and go back to $B_{s}=0$ at $W\approx10.5$ [shown in Fig.~\ref{fig4}(f)]. It is obvious that the plateaux of spin Bott index can match well with the plateaux of the conductance. However, some similar quantized conductance and spin Bott index plateaux with the range of $W_{l}\approx(6.5, 7, 8)$ to $W\approx10$ also emerge in the crystalline system, as shown in Figs.~\ref{fig4}(d) and~\ref{fig4}(h). It is found that the disorder-induced topological phase transitions from normal insulator to TAI and then to Anderson insulator occur in both the quasicrystalline and the crystalline systems. However, for the same model parameters, the disorder-induced quantized plateau in the quasicrystalline system appears in smaller disorder strength and vanishes in stronger disorder strength than that in the crystalline system. Thus, the disorder-induced QSH insulator phases are qualitatively similar in the quasicrystalline system and the crystalline system, but quantitatively different.

\begin{figure}[t]
	\includegraphics[width=8.5cm]{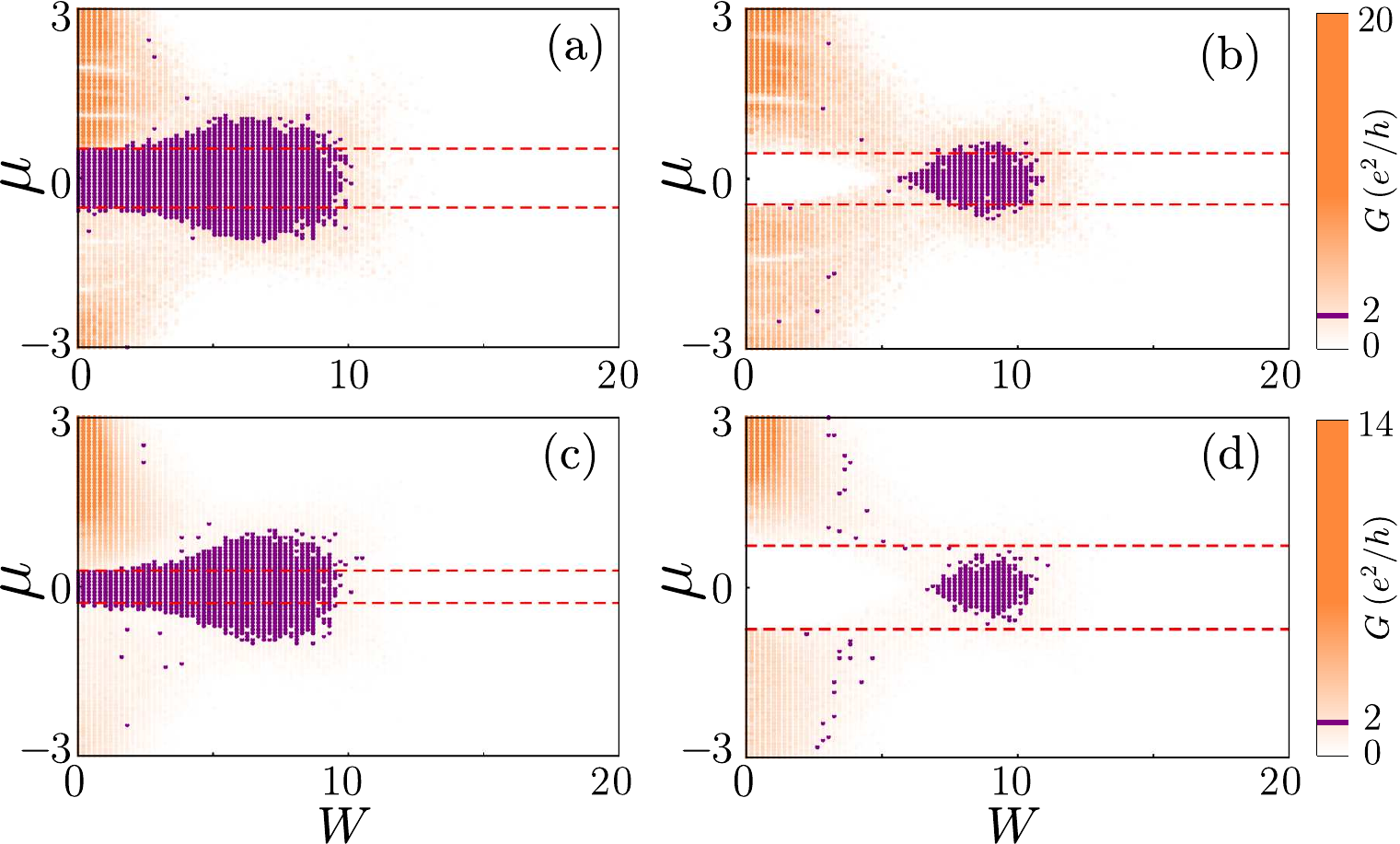} \caption{Topological phase diagram in ($\mu$, $W$) space of the two systems obtained by calculating the two-terminal conductance. (a) The quasicrystalline system with $M=1$, (b) the quasicrystalline system with $M=2$, (c) the crystalline system with $M=1$, and (d) the crystalline system with $M=2$. The purple region denotes the topologically nontrivial phase ($G=2e^2/h$), and the white region denotes the topologically trivial phase ($G=0$). We set $t_2=1$. The conductance is calculated with the device size being $L_{x}\times L_{y}=200r_1\times100r_1$.}%
\label{fig5}
\end{figure}

Figure~\ref{fig5} shows the topological phase diagrams in ($\mu$, $W$) space of the quasicrystalline system and the crystalline system with different parameters obtained by the two-terminal conductance. The color map shows the values of conductance $G$. Each point in Fig.~\ref{fig5} is obtained from a single distribution of disorder, which is sufficient to determine the quantized conductance region. The purple region represents the topological nontrivial phase with a quantized conductance $G=2e^2/h$, the white region represents the topological trivial phase with $G=0$, and the orange region corresponds to the conductance of the bulk. The difference of the two lattice construction may lead to the difference of the maximum bulk conductance. The region between the two red dashed lines represents the bulk energy gap of the system. It is obvious that the TAI phase occurs mainly in the energy gap and the the disorder-induced topological nontrivial phase region in the Ammann-Beenker tiling quasicrystal is larger and more stable than that in the snub-square crystalline lattice, as shown in Figs.~\ref{fig5}(b) and \ref{fig5}(d).

\begin{figure}[t]
	\includegraphics[width=8cm]{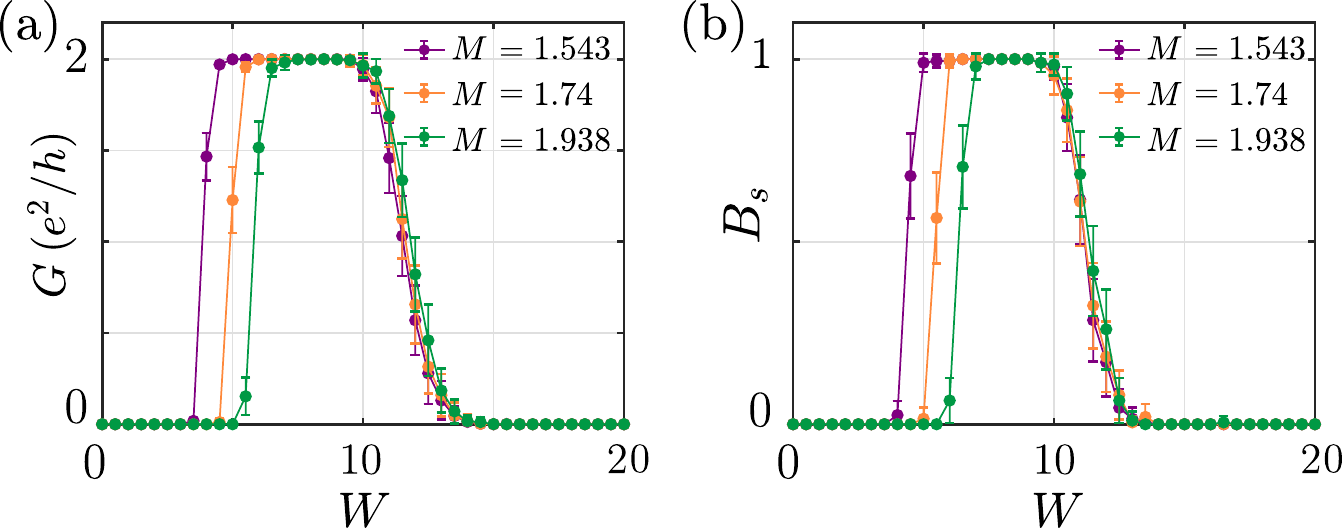} \caption {(a) The two-terminal conductance $G$ in crystalline system as a function of the disorder strength $W$ with different Dirac mass $M=1.543$, $1.74$, and $1.938$, respectively. (b) The spin Bott index $B_{s}$ in crystalline system as a function of the disorder strength $W$ with different Dirac mass $M=1.543$, $1.74$, and $1.938$, respectively. The chemical potential is taken as $\mu=0$. The error bars show the standard deviation of the conductance for $1000$ samples. We set $t_2=1$. The conductance is calculated with the device size being $L_{x}\times L_{y}=200r_1\times 100r_1$. The spin Bott index is calculated with a square sample containing 1393 sites for quasicrystalline system and 1444 sites for crystalline system, respectively.}%
\label{fig6}
\end{figure}

In fact, due to the different lattice constructions, the bulk energy gaps are different for the two systems with the same model parameters. We will take some specific model parameters to ensure that the two systems have the same bulk energy gaps in the clean limit, and then study the effects of disorder in this case by calculating the two-terminal conductance and spin Bott index. Figure ~\ref{fig6} shows the two-terminal conductance and the spin Bott index as a function of the disorder strength $W$ with $M_{l}\approx(1.543, 1.740, 1.938)$ in crystal. The corresponding bulk energy gaps are $E_{l}\approx(0.510, 0.903, 1.299)$ in the clean limit, which are very closed to the energy gaps $E_{l}\approx(0.511, 0.905, 1.299)$ with $M_{l}=(1.8, 2.0, 2.2)$ in quasicrystal. It is found that the conductance and spin Bott index plateaux arise at $W_{l}\approx(5, 6, 7 )$ and vanish at $W\approx9.5$, as shown in Figs.~\ref{fig6}(a) and~\ref{fig6}(b). Here, the points where the conductance and spin Bott index plateaus appear are consistent with that in Figs.~\ref{fig4}(a) and ~\ref{fig4}(b). Thus, it is indicated that the response in the two systems of disorder is roughly same when the initial bulk energy gaps of the quasicrystalline system and the crystalline system are the same.

In order to better understand the origin of the TAI phase, we calculate the nonequilibrium local current between neighboring sites $i$ and $j$ from the formula \cite{PhysRevB.80.165316}
\begin{equation}
J_{\mathbf{{i}\rightarrow{j}}}=\frac{2e^{2}}{h}\text{Im} \left[  \sum
_{a,b}{H_{\mathbf{i} a,\mathbf{j}b}G^{n}_{\mathbf{j}b
			,\mathbf{i}a}}(E_{F})\right]  \left(  V_{L}-V_{R} \right)  \text{,}%
\label{Current}
\end{equation}
where $G^{n}(E_{F})=G^{r}(E_{F})\Gamma_{L} (E_{F})G^{a}(E_{F})$ is the electron correlation function. $a$ and $b$ denote the state indices. To calculate the local current distribution, a small external bias $V=V_L-V_R$ is applied longitudinally between the two terminals, where $V_L$ and $V_R$ describe the voltages of the left and right leads. The small bias voltage $V$ is fixed to be $0.001$.

\begin{figure}[t]
	\includegraphics[width=7.5cm]{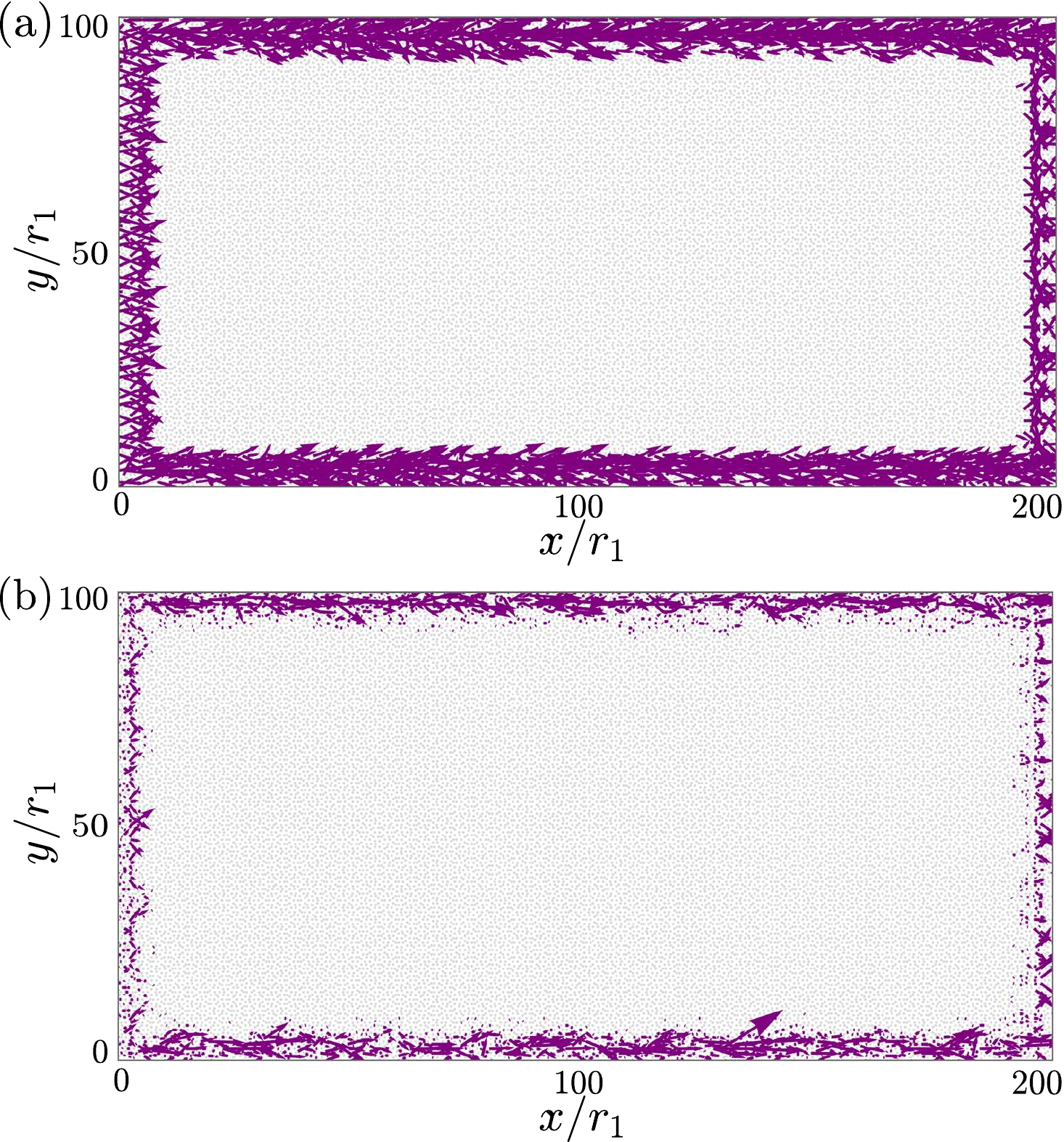} \caption {The averaged nonequilibrium local current distribution of the disorder-induced QSH insulator phases of (a) the quasicrystalline system and (b) the crystalline system. The arrow size and direction represent the local current strength and direction, respectively. We set $M=2$, $t_2=1$, and $W=8$.}%
\label{fig7}
\end{figure}

In Fig.~\ref{fig7}, we show the averaged current distribution of the disorder-induced QSH insulator state. Distinctly, in both the quasicrystalline system and the crystalline system, the currents tend to be zero in the bulk, and are localized at the boundaries of the samples without much scattering. The edge transport is a strong evidence of the quantized conductance, which is contributed by the topological Anderson insulator phase. The numerical results are consistent with the above results of the two terminal conductance and the spin Bott index.

\section{Conclusion and discussion}
\label{Conclusion}
In this paper, based on the spin Bott index and the two-terminal conductance method, we investigate the disorder-induced topological phase transitions of the Ammann-Beenker tiling quasicrystal and a periodic snub-square crystal. The two lattices have the same square and rhombus tiles. The numerical calculating results of the spin Bott index and the two-terminal conductance of the two systems share qualitative similarities. First, the QSH insulator phase in the two systems are robust against weak disorder and destroyed by strong disorder. Second, the topological phase transitions from normal insulator to TAI characterized by a quantized conductance plateau and a nonzero spin Bott index of the two systems occur at a certain range of disorder strength. Some quantitative differences are also observed, such as the conductance plateau induced by disorder in the Ammann-Beenker tiling quasicrystal is wider than that in the snub-square crystal. Furthermore, we present the averaged nonequilibrium local current distribution to further identify the disorder-induced topological nontrivial phases of the two systems. In summary, we find that the disorder-induced topological phase transitions whether from QSH insulator to normal insulator or from normal insulator to TAI are independent of the lattice structure by comparing the results in the Ammann-Beenker tiling quasicrystal with those of the periodic snub-square crystal.

It is noted that the classification of quasicrystals in two dimensions is a very important problem, and has been investigated in previous works \cite{PhysRevLett.58.2099,RevModPhys.69.1181,PhysRevLett.93.045501}. The problem of classifying 2D lattices with $N$-fold rotational symmetry for arbitrary (noncrystallographic) even $N$ is equivalent to a much-studied problem in algebraic number theory \cite{PhysRevLett.58.2099}. Mermin \emph{et al.} found that except for 29 even number $N$ there are two or more distinct lattices, and the smallest $N$ for which there is more than a single lattice is $N$=46 \cite{PhysRevLett.58.2099}. In the previous work, the TAI was theoretically investigated in a fivefold Penrose tiling quasicrystal \cite{PhysRevB.100.115311}. Here our study further confirmed that the TAI can occur in an eightfold Ammann-Beenker tiling quasicrystal. In addition, we also calculated the two-terminal conductance of a twelvefold Stampfli-tiling quasicrystal with disorder [the model Hamiltonian is the same as Eqs. (1)-(3)]. It is found that the QSH and the TAI phase can also occur in a twelvefold Stampfli-tiling quasicrystal (further details will be given in the future work). We conjecture that the appearance of the TAI phase in 2D quasicrystals would be a general phenomenon, just the values of parameters where the topological phase transition occurs depend on the quasicrystal lattice structure properties. The TAI phase of the other N-fold quasicrystals will be furtherly investigated in the future, and we predict that the same physics will be implemented in these lattices.

Recently, TAIs and quasicrystals have been successfully implemented in experiments based on photonic platform \cite{PhysRevX.6.011016,stutzer2018photonic,levi2011disorder,PhysRevLett.110.076403,vardeny2013optics,PhysRevLett.122.110404,Steurer_2007}. Inspired by the recent experiments, we propose an experimental setup to construct an Ammann-Beenker tiling quasicrystalline lattice and a periodic snub-square crystalline lattice in an array of helical evanescently coupled waveguides \cite{PhysRevX.6.011016,PhysRevLett.122.110404}, and add on-site disorder in the form of random variations in the refractive index of the waveguides \cite{stutzer2018photonic}. Consequently, we believe that the experimental realization of the Ammann-Beenker tiling quasicrystalline lattice and the periodic snub-square crystalline lattice with disorder is promising in the
above experimental schemes, especially in the photonic systems.

\section*{Acknowledgments}

 B.Z. was supported by the NSFC (under Grant No. 12074107) and the program of outstanding young and middle-aged scientific and technological innovation team of colleges and universities in Hubei Province (under Grant No. T2020001). D.-H.X. was supported by the NSFC (under Grants No. 12074108 and No. 11704106). D.-H.X. also acknowledges the financial support of the Chutian Scholars Program in Hubei Province. R.C. was supported by the Project funded by China Postdoctoral Science Foundation (under Grant No. 2019M661678).

\end{document}